\documentstyle[12pt]{article}

\def\thebibliography#1{\bigskip\section*{\large
\bf References\\}\list
  {[\arabic{enumi}]}{\settowidth\labelwidth{#1}\leftmargin\labelwidth
    \advance\leftmargin\labelsep
    \usecounter{enumi}}
    \def\newblock{\hskip .11em plus .33em minus .07em}
    \sloppy\clubpenalty4000\widowpenalty4000
    \sfcode`\.=1000\relax}

\def\op#1{\mathop{{\it\fam0} #1}\limits}

\newcommand{\Ker}{{\rm Ker\,}}

\newcommand{\nm}[1]{\mid {#1}\mid}

\newcommand{\beq}{\begin{equation}}
\newcommand{\eeq}{\end{equation}}
\newcommand{\ben}{\begin{eqnarray}}
\newcommand{\een}{\end{eqnarray}}
\newcommand{\be}{\begin{eqnarray*}}
\newcommand{\ee}{\end{eqnarray*}}
\newcommand{\bea}{\begin{eqalph}}
\newcommand{\eea}{\end{eqalph}}

\newcommand{\cL}{{\cal L}}

\newcommand{\bR}{{\bf R}}
\newcommand{\bC}{{\bf C}}
\newcommand{\bZ}{{\bf Z}}

\newcommand{\rs}{{\rm s}}

\newcommand{\al}{\alpha}
\newcommand{\bt}{\beta}
\newcommand{\dl}{\delta}
\newcommand{\la}{\lambda}

\newcommand{\Om}{\Omega}
\newcommand{\m}{\mu}
\newcommand{\n}{\nu}
\newcommand{\g}{\gamma}

\newcommand{\ve}{\varepsilon}
\newcommand{\th}{\theta}

\newcommand{\si}{\sigma}
\newcommand{\Si}{\Sigma}

\newcommand{\wt}{\widetilde}

\newcommand{\dr}{\partial}

\newcommand{\ar}{\op\longrightarrow}

\newcommand{\ot}{\otimes}

\newenvironment{eqalph}{\stepcounter{equation}
\setcounter{equationa}{\value{equation}}
\setcounter{equation}{0}

\begin{eqnarray}}{\end{eqnarray}\setcounter{equation}{\value{equationa}}}

\newcommand{\mar}[1]{}

\begin{document}
\hbox{}

{\parindent=0pt

{\large \bf CLASSICAL GAUGE THEORY OF GRAVITY}\footnote{{\it
Theoretical and Mathematical Physics} {\bf 132} (2002) 1163-1171;
translated from {\it Teoreticheskaya i Matematicheskaya Fizika} {\bf
132} (2002) 318-328}
\bigskip 

{\bf G. A. Sardanashvily}

\medskip

\begin{small}

Department of Theoretical Physics, Moscow State University, 117234
Moscow, Russia

E-mail: sard@grav.phys.msu.su

URL: http://webcenter.ru/$\sim$sardan/
\bigskip

{\it The classical theory of gravity is formulated as a gauge theory on
a frame bundle with spontaneous symmetry of breaking caused by the
existence of Dirac fermionic fields. The pseudo-Riemannian metric
(tetrad field) is the corresponding Higgs field. We consider two
variants of this theory. In the first variant, gravity is represented
by the pseudo-Riemannian metric as in general relativity theory; in the
second variant, it is represented by the effective metric as in the
Logunov relativistic theory of gravity. The configuration space, Dirac
operator, and Lagrangians are constructed for both variants.
}

\bigskip

{\bf Key words:} gravity, gauge field, Higgs field, spinor field

\end{small}
}


\section{Introduction}

The first gauge model of gravity [1] was built just two years after
birth of the gauge theory itself. But the initial attempts to construct
the gauge theory of gravity by analogy with the gauge models of
internal symmetries encountered a serious difficulty -- establishing
the gauge status of the metric (tetrad) field and choosing the gauge
group. In [1], gravity was described in the framework of the gauge
model of the Lorentz group, but tetrad fields were introduced
arbitrarily. To eliminate this drawback, representing tetrad fields as
gauge fields of the translation group was attempted [2]-[9], but to no
effect. At first, the canonical lift of vector fields on a manifold
$X$ to the tangent bundle $TX$ [2], [3], which is actually the
generator of general covariant transformations of $TX$, was considered
as generators of the translation gauge group. A
horizontal lift of vector fields on $X$ using a linear connection on
$TX\to X$ was proposed for the same role [4]. The tetrad field was
later identified with the translation part of an affine connection on
the tangent bundle $TX\to X$. Any such connection is a sum of a linear
connection and a tangent-valued form
\mar{mos035}\beq
\Om=\Om^\al_\la(x) dx^\la\ot\dr_\al, \label{mos035}
\eeq
but this form and the tetrad field are two different mathematical
objects [10], [11]. We note that form (1) keeps appearing in one of the
variants of the gauge theory of gravity as a nonholonomic frame [12].
At the same time, it turned out that translational connections describe
distortion in the gauge theory of dislocations [13], [14] and in the
analogous gauge model of the "fifth force" [11], [15].

Difficulties of constructing the gauge theory of gravity by analogy
with the gauge theory of internal symmetries resulted from the gauge
transformations in these theories belonging to different classes. In
the case of internal symmetries, the gauge transformations are just
vertical automorphisms of the principal bundle $P\to X$ leaving its base
$X$ fixed. On the other hand, gravity theory is built on the principal
bundle $LX$ of the tangent frames to $X$ and tensor bundles to $X$
associated with it. They 
belong to the category of bundles $T\to X$ for which diffeomorphisms
$f$ of the base $X$ canonically generate automorphisms
$\wt f$ of the bundle $T$. These automorphisms are called holonomic
automorphisms or general covariant transformations.
All gravitational Lagrangians are constructed to be invariant under
general covariant transformations, which are therefore gauge
transformations of gravity theory. Nonholonomic (e.g., vertical)
automorphisms of the principal bundle $LX$ can also be considered, but
most gravitational Lagrangians are not left invariant under these
transformations. 

In the persistent attempts to represent the terad field as a gauge
field, the presence of Higgs fields in addition to the matter and gauge
fields in the gauge theory with spontaneous symmetry breaking was
overlooked. Spontaneous symmetry breaking is a quantum effect when the
vacuum is not invariant under the transformation group. In terms of the
classical gauge theory, spontaneous symmetry breaking occurs if the
structure group $G$ of the principal bundle $P\to X$ is reducible to a
closed subgroup $H$, i.e., there exists a principal subbundle of the
bundle $P$ with the structure group $H$ [10], [16]-[18]. Such a reduction
occurs if and only if the quotient bundle $P/H\to X$ admits a global
section. Moreover, there exists a one-to-one correspondence between
reduced subbundles $P^h\subset P$ with the structure group $H$ and
global sections $h$ of the quotient bundle $P/H\to X$ [19]. The latter
are treated as classical Higgs fields.

The idea of the pseudo-Riemannian metric as a Higgs-Goldstone field
appeared while constructing nonlinear (induced) representations of the
group $GL(4,\bR)$, of which the Lorentz group is a Cartan subgroup
[20], [21]. When the gauge theory was formulated in terms of bundles,
the very definition of the pseudo-Riemannian metric on a manifold $X$
as a global section of the quotient bundle  
\mar{b3203}\beq
\Si_{PR}= LX/SO(1,3)\to X\label{b3203}
\eeq
led to its physical interpretation as a Higgs field responsible for
the spontaneous space-time symmetry breaking [10, [22], [23]. The
geometric principle of equivalence postulating the existence of a
reference frame in which Lorentz invariants are defined on the whole
space-time manifold is the theoretical justification of such symmetry
breaking [10]. In terms of bundles, this means that the structure group
$GL(4,\bR)$ of the frame bundle and the bundles associated with it is
reduced to the Lorentz group.

The physical reason for space-time symmetry breaking is the existence
of Dirac fermionic matter, whose symmetry group is the universal
two-sheeted covering $L_s=SL(2,\bC)$ of the proper Lorentz group
$L=SO^\uparrow(1,3)$. The Dirac spin structure on a space-time manifold
$X$ is defined as a pair
$(P^h, z_\rs)$ of a principal bundle $P^h\to X$ with the structure
group $L_s$ and the fiberwise morphism
\mar{b3246}\beq
z_s: P^h \to LX \label{b3246}
\eeq
to the frame bundle $LX$ [24], [25]. Any such morphism can be factored
using a morphism
\mar{mos250}\beq
z_h: P^h \to L^hX, \label{mos250}
\eeq
where $L^hX$ is a principal subbundle of the frame bundle $LX$
with the structure group being the Lorentz group $L$. 

We assume in what follows that $X$ is a four-dimensional oriented
(simply connected, smooth, separate, locally compact, countably
infinite, i.e., paracompact) manifold with a coordinate atlas
$\{(U_\zeta,x^\la)\}$ and that $LX$ is the bundle of oriented tangent
frames to $X$ with the structure group $GL_4=GL^+(4,\bR)$. Then there
exists a one-to-one correspondence between the abovementioned reduced
bundles $L^hX$ with the structure group $L$ and global sections
$h$ of the quotient bundle
\mar{5.15}\beq
\Si=LX/L\to X \label{5.15}
\eeq
with a typical fiber $GL_4/L$. In other words, the restriction of the
bundle $LX\to \Si$ to $h(X)\subset \Si$ is isomorphic to
$L^hX$. Bundle (5) is a two-sheeted covering of the bundle $\Si_{PR}$ 
given by (2). Its global section $h$ is called the tetrad field. It is
represented by a collection of local sections $h^a_\m(x)$ of the
reduced subbundle $L^hX$, which are called tetrad functions. Any tetrad
field uniqiely determines a pseudo-Riemannian metric on $X$ such that
the well-known identity 
\mar{mos175}\beq
g_{\mu\nu}=h^a_\mu h^b_\nu\eta_{ab}, \label{mos175}
\eeq
holds, where $\eta$ is the Minkowski metric. A Dirac spin structure on
the space-time manifold thus uniquely determines a pseudo-Riemannian
metric on it.

The existnece of a Dirac spin structure also imposes severe
restrictions on the topology of the space-time manifold $X$
[26], [27]. Because compact space-time does not satisfy any casuality
principle, we assume in what follows that the manifold $X$ is not
compact. Then it is parallelizable, i.e., the frame bundle $LX\to X$ is
trivial. In this case, all spin structures $P^h$ on $X$ are mutually
isomorphic although this isomorphism is not canonical.
 
Let $S^h\to X$ denote the spinor bundle associated with the principal
bundle $P^h$. Its sections describe Dirac fermionic fields in the
presence of a tetrad field $h$. The problem is that in the presence of
different tetrad fields, the fermionoc fields are represented by
sections of different bundles with no canonical isomorphisms between
them. Moreover, to define the Dirac operator on the sections of the
bundle $S^h$,
we must specify a representation
\mar{gg1}\beq
dx^\la\mapsto \g_h(dx^\la)=h^\la_a(x)\g^a \label{gg1}
\eeq
of coframes $\{dx^\la\}$ tangent to the space-time manifold $X$ by
Dirac matrices. But these representations are not equivalent for
different tetrad fields $h$. The Higgs nature of the classical tetrad
field manifests itself here. A Dirac field on the space-time manifold
can only be considered paired with a definite tetrad field.

In particular, because morphism (3) is factored using morphism (4), the
whole existing fermionic matter represented by sections of some spinor
bundle over $X$ uniquely fixes the Lorentz structure $L^hX$ and tetrad
field $h$
(pseudo-Riemannian metric $g$) on $X$, this being independent of the
fermionic field dynamics. Therefore, this tetrad field
(pseudo-Riemannian metric) is not determined by any differential
equations and is a background field. Consequently, it cannot be
idetified with the gravitational field, and a background geometry is
present in the gravity theory. This considerations motivate us to
regard two variants of the gauge theory of gravity.

In the first variant [11], [28]-[31], the gravitational field is
identified with the tetrad field, and we consider the composite bundle
\mar{y2}\beq
S\to \Si \to X, \label{y2}
\eeq
where $S\to\Si$ is the spinor bundle associated with the principal
bundle $LX\to\Si$ with the structure group being the Lorentz group $L$.
The idea of this construction is that for any section $h$ of the tetrad
bundle $\Si\to X$, the restriction of the bundle $S\to\Si$ to
$h(X)\subset \Si$ is isomorphic to the spinor bundle 
$S^h\to X$, whose sections are fermionic fields in the presence of a
tetrad gravitational field $h$. Therefore, sections of bundle (8)
describe the totality of fermionic and gravitational fields.

The second variant [29], [32] uses the fact that if 
$Q$ is a group bundle associated with $LX$, then there exists a morphism
\mar{hhh}\beq
\rho:Q\times \Si\to \Si \label{hhh}
\eeq
that assigns to any tetrad field $h$ a background tetrad field $h_0$
and a section $q$ of the bundle $Q\to X$, 
the latter section being identified with the gravitational field. In
this approach, Dirac fermionic fields are described by sections of the
bundle $S^{h_0}\to X$, and their dynamics in the presence of
gravitational field $q$ appears as motion in the effective tetrad field
$\wt h$.

We define the configuration space, Dirac operator, and Lagrangians for
both variants of the gauge theory of gravity.

\section{Lorentz structure}

As already stated, the theory of gravity is built as a gauge theory on
the principal bundle $LX\to X$ of oriented frames tangent to the
space-time $X$ with the structure group 
$GL_4$. We say a few words about this bundle. Any element of it can be
represented in the form $H_a=H^\m_a\dr_\m$, where 
$\{\dr_\m\}$ is the set of holonomic bases of the tangent bundle $TX$ and
$H^\m_a$ are the matrix elements of the representation of the group
$GL_4$ in $\bR^4$. The latter serve as coordinates on the bundle
$LX$ with the transition functions
\be
H'^\m_a=\frac{\dr x'^\m}{\dr x^\la}H^\la_a.
\ee
In these coordinates, the right action of the structure group
$GL_4$ on $LX$ becomes
\mar{ccc}\beq
R_g: H^\m_a\mapsto H^\m_bg^b{}_a, \qquad g\in GL_4. \label{ccc}
\eeq
The frame bundle $LX$ is equipped with a canonical $\bR^4$-valued 1-form 
\mar{b3133'}\beq
\th_{LX} = H^a_\m dx^\m\ot t_a, \label{b3133'}
\eeq
where $\{t_a\}$ is a fixed basis for $\bR^4$ and $H^a_\m$ are elements
of the inverse matrix.

As in any gauge theory, connections on the principal frame bundle $LX$
are the gauge fields in the gravity theory. They are in one-to-one
correspondence with linear connections
$K$ on the tangent bundle $TX$ (or just on $X$). In the holonomic
coordinates $(x^\la,\dot x^\la)$ on $TX$, these connections are given
by the tangent-valued forms
\mar{B}\beq
K= dx^\la\otimes (\dr_\la +K_\la{}^\m{}_\n \dot x^\n
\dot\dr_\m) \label{B}
\eeq
and are represented by sections of the quotient bundle 
\mar{015}\beq
C=J^1LX/GL_4\to X, \label{015}
\eeq
where $J^1LX$ is the manifold of first-order jets of sections of the
bundle $LX\to X$
[31]. The bundle of connections $C$ is equipped with the coordinates
$(x^\la, k_\la{}^\nu{}_\al)$ such that the coordinates
$k_\la{}^\nu{}_\al\circ K=K_\la{}^\nu{}_\al$ of any section $K$
are the coefficients of corresponding linear connection (12). We stress
that the bundle of connections $C$ given by (13) is not associated with
the frame bundle but also admits general covariant transformations.

As already mentioned, the metric (tetrad) field is introduced in the
gauge theory of gravity by specifying the Lorentz structure. The subbundle
$L^hX$ of the frame bundle $LX$ with the structure group
$L$, where $h$ is a global section of quotient bundle (5), is called
the Lorentz structure on the space-time manifold $X$.

The tetrad field $h$ determines the atlas 
$\Psi^h=\{(U_\zeta,z_\zeta^h)\}$ of the frame bundle 
$LX$ such that local sections
$z_\zeta^h$ of the bundle $LX$ take values in the Lorentz subbundle
$L^hX$ and have Lorentz transition functions. They are called the
tetrad functions and have the form
$h^\m_a =H^\m{}_a\circ z^h_\zeta$.
Tetrad functions induce the local tetrad form 
\be
z_\zeta^{h*}\th_{LX}=h_\la^a dx^\la\ot
t_a
\ee
on $X$, where $\th_{LX}$ is canonical form (11) on
$LX$. This tetrad form in turn determines tetrad coframes
\be
h^a =h^a_\m(x)dx^\m, \qquad x\in U_\zeta, 
\ee
on the cotangent bundle $T^*X$ of $X$. In particular, identity (6)
becomes
$g=\eta_{ab}h^a\ot h^b$.
The metric $g$ is thus reduced to the Minkowski metric with respect to
the Lorentz atlas $\Psi^h$ and is an example of Lorentz invariants in
the geometric equivalence principle.

We now consider gauge fields in the presence of the Lorentz structure.
Connections on the Lorentz bundle $L^hX$ have the form
\mar{b3205}\beq
A_h=dx^\la\ot(\dr_\la + \frac12A_\la{}^{ab} \ve_{ab}),
\label{b3205}
\eeq
where $\ve_{ab}{}^c{}_d= \eta_{ad}\dl^c_b- \eta_{bd}\dl^c_a$ are the
generators of Lorentz group. We call them Lorentz connections. Taking
the property of equivariance with respect to right action (10) of the
structure group on the principal bundle into account, we can extend any
Lorentz connection on $L^hX$ to a connection on the frame bundle
$LX$ thus determining a linear connection $K$ (see (12)) on $X$ with
coefficients  
\mar{mos190}\beq
K_\la{}^\m{}_\nu = h^k_\nu\dr_\la h^\m_k + \eta_{ka}h^\m_b h^k_\nu
A_\la{}^{ab}. \label{mos190}
\eeq
We can formulate the converse statement. According to the known theorem
[19], a linear connection on $X$
is a Lorentz connection if and only if its holonomy group reduces to
the Lorentz group. At the same time, it can be shown [29], [31] that
on any Lorentz subbundle $L^hX$, any linear connection $K$ given by
(12) determines a Lorentz connection $A_h$ given by (14) with the
coefficients 
\mar{K102}\beq
A_\la{}^{ab} =\frac12 (\eta^{kb}h^a_\m-\eta^{ka}h^b_\m)(\dr_\la h^\m_k -
 h^\nu_k K_\la{}^\m{}_\nu). \label{K102}
\eeq
This allows using general connections independent of any spin structure
in a gravity theory with fermionic fields.

\section{Universal spin structure}

Because the first homotopic group of the group space $GL_4$ 
equals $\bZ_2$,
there exists a universal covering group $\wt{GL}_4$ of the group $GL_4$
such that the diagram
\be
\begin{array}{ccc}
 \wt{GL}_4 & \longrightarrow &  GL_4 \\
 \put(0,-10){\vector(0,1){20}} & 
& \put(0,-10){\vector(0,1){20}}  \\
L_s & \ar^{z_L} &  L
\end{array} 
\ee
is commutative. The group $\wt{GL}_4$ has a spinor representation whose
elements, called "world" spinors, were proposed for describing fermions
in gravity theory [12]. But this representation is
infinite-dimensional. We therefore choose another way.

We consider the two-sheeted covering $\wt{LX}\to X$ of the frame bunde 
$LX$, which is a principal bundle with the structure group 
$\wt{GL}_4$ [25], [33], [34]. Because $\Si=\wt{LX}/L_\rs$, the bundle
\mar{mos265}\beq
\wt{LX}\to\Si \label{mos265}
\eeq
is a principal bundle with the structure group $L_s$. Therefore, the
commutative diagram 
\mar{b3250}\beq
\begin{array}{rcl}
 \wt{LX}  & \op\longrightarrow^{\wt z} &  LX \\
  & \searrow  \swarrow & \\ 
 & {\Si} &  
\end{array} \label{b3250}
\eeq
is valid, which gives the spin structure on the tetrad bundle $\Si$.
This spin structure is unique and has the following property [29]-[31].
For any global section $h$ of the bundle $\Si\to X$ given by (5), the
restriction of principal bundle (17) to $h(X)\subset \Si$ is isomorphic
to the principal bundle $P^h$ given by (3) with the structure group
$L_s$. Therefore, spin structure (18) is said to be universal.

We consider the spinor bundle $S\to\Si$ associated with (17). Similarly
to the preceding, if $h$ is a section of the bundle $\Si\to
X$, then the restriction $h^*S$ of the bundle $S\to\Si$ to $h(X)\subset
\Si$ is a subbundle of the composite bundle $S\to X$ given by (8) and
is isomorphic to the spinor bundle $S^h$ associated with $P^h$. We note
that $S\to X$ is not a spinor bundle. In particular, holonomic
automorphisms of the frame bundle $LX$ unambigously extend to $\wt{LX}$ 
and induce general covariant transformations of the bundle $S\to X$
[28]-[31]. 

We construct the Dirac operator on the spinor bundle $S\to\Si$ such
that its restriction to the subbundle $S^h$ of the bundle $S\to X$
reproduces the Dirac operator of fermionic fields in the presence of a
tetrad field $h$ and a general linear connection $K$ on $X$.
We recall that connections on the spinor bundle $S^h\to X$ are
associated with connections on the principal bundle $P^h$ and are in
one-to-one correspondence with Lorentz connections on $L^hX$.
It follows that any Lorentz connection $A_h$ given by (14) determines
the spinor connection
\mar{yyy}\beq
A_h=dx^\la\ot(\dr_\la +
A^{ab}{}_\la L_{ab}{}^A{}_By^B \dr_A), \qquad
L_{ab}=\frac{1}{4}[\g_a,\g_b], \label{yyy} 
\eeq 
on the bundle $S^h$ equipped with the coordinates $(x^\la,y^A)$ 
relative to the Lorentz atlas $\Psi^h=\{h^a_\la\}$ 
of the bundle $L^hX$ extended to $P^h$. Consequently, an arbitrary
linear connection $K$ given by (12) on 
$X$ induces a Lorentz connection on $L^hX$ with coefficients (16) and
correspondingly a spinor connection  
\mar{b3212}\beq
K_h=dx^\la\ot[\dr_\la +\frac14 (\eta^{kb}h^a_\m-\eta^{ka}h^b_\m)(\dr_\la
h^\m_k - h^\nu_k K_\la{}^\m{}_\nu)L_{ab}{}^A{}_B y^B\dr_A] \label{b3212}
\eeq
on $S^h$ [28], [31], [35]. Spinor connection (19) determines the
covariant differential 
\mar{ppp}\beq
D: J^1S^h\to T^*X\ot S^h,\qquad
D=(y^A_\la-A^{ab}{}_\la L_{ab}{}^A{}_By^B)dx^\la\ot\dr_A, \label{ppp}
\eeq
on $S^h$, where $J^1S^h$ with coordinates $(x^\la, y^A,y^A_\la)$ is the
manifold of jets of sections of the fibre bundle $S^h\to X$.
On the bundle $S^h$, the covariant differential $D$ given by (21) in
composition with representation (7) determines the Dirac operator
\mar{l13}\ben
&& \Delta_h=\g_h\circ D: J^1S^h\to T^*X\ot S^h\to S^h, \label{l13}\\
&& y^A\circ\Delta_h=h^\la_a \g^{aA}{}_B(y^B_\la- \frac12 A^{ab}{}_\la
L_{ab}{}^A{}_By^B) \nonumber
\een
of fermionic fields in the presence of a background tetrad field $h$
and a general linear connection $K$.

We now turn to the bundle $S\to \Si$ and spinor bundle $Y$ over the
product $\Si\times C$ induced from it, where $C$ is the bundle of
general linear connections (13). The bundle $Y$ is equipped with the
coordinates $(x^\la,y^A,\si^a_\mu,k_\la{}^\mu{}_\nu)$, where
$(x^\la,\si^a_\mu)$ are coordinates on the tetrad bundle $\Si$ such
that the coordinates $\si^a_\mu\circ h$ of any section $h$ are tetrad
functions $h^a_\m$. 
Œ­®£®®¡à §The manifold of jets $J^1Y$ of the bundle $Y\to \Si\times
C\to X$ is the configuration space of the complete system of fermionic,
tetrad, and gauge fields in the gauge theory of gravity where the
tetrad field and corresponding pseudo-Riemannian metric are identified
with the gravitational field. Dirac operator on this configuration
space has the form [28]-[31]
\mar{b3264}\ben
&& \Delta_Y=\g_\Si\circ\wt D:J^1Y\to T^*X\op\ot_\Si S\to S,
\nonumber\\
&& y^B\circ\Delta_Y=\si^\la_a\g^{aB}{}_A[y^A_\la-  \frac14(\eta^{kb}\si^a_\m
-\eta^{ka}\si^b_\m)(\si^\m_{\la k} -\si^\nu_k
k_\la{}^\m{}_\nu)L_{ab}{}^A{}_By^B]. \label{b3264}
\een
It satisfies the requirement stated above and coincides with Dirac
operator (22) when restricted to $h(X)\times
K(X)\subset \Si\times C$ for a background tetrad field $h$ and a linear
connection $K$. 
 
The Lagrangian of the gauge theory of gravity on the configuration
space $J^1Y$ can be chosen in the form of a sum
\mar{y12}\beq
L=L_D + L_{AM}  \label{y12}
\eeq
of the Dirac Lagrangian
\mar{b3265}\ben
&& L_D=\{\frac{i}{2}\si^\la_q[y^+_A(\g^0\g^q)^A{}_B(y^B_\la-
\frac14(\eta^{kb}\si^a_\m
-\eta^{ka}\si^b_\m)(\si^\m_{\la k} -\si^\nu_k
k_\la{}^\m{}_\nu)L_{ab}{}^B{}_Cy^C)- \nonumber\\
&& \qquad (y^+_{\la A} -
\frac14(\eta^{kb}\si^a_\m
-\eta^{ka}\si^b_\m)(\si^\m_{\la k} -\si^\nu_k
k_\la{}^\m{}_\nu)y^+_C L^+_{ab}{}^C{}_A(\g^0\g^q)^A{}_By^B]- \label{b3265}\\ 
&&\qquad  my^+_A(\g^0)^A{}_By^B\}\sqrt{\nm\si}, \qquad \si=\det(\si_{\m\n}),
\nonumber
\een
and the Lagrangian 
\be
L_{AM}(R_{\m\la}{}^\al{}_\bt,\si^{\m\nu}), \qquad
\si^{\m\n}=\si^\m_a\si^\nu_b\eta^{ab},
\ee
of the affine-metric theory expressed via the curvature tensor of the
linear connection 
\mar{www}\beq
R_{\la\m}{}^\al{}_\bt = k_{\la\m}{}^\al{}_\bt - k_{\m\la}{}^\al{}_\bt
+ k_\m{}^\al{}_\ve k_\la{}^\ve{}_\bt
-k_\la{}^\al{}_\ve k_\m{}^\ve{}_\bt. \label{www}
\eeq
It is easy to see that
\be
\frac{\dr\cL_D}{\dr k_\la{}^\m{}_\nu} + 
\frac{\dr\cL_D}{\dr k_\nu{}^\m{}_\la} =0.
\ee
Therefore, Dirac Lagrangian (25) depends only on the torsion tensor
$k_\la{}^\m{}_\nu-k_\nu{}^\m{}_\la$ of the linear connection. Moreover,
full Lagrangian (24) is invariant under nonholonomic gauge transformations
\be
k_\la{}^\m{}_\nu \to k_\la{}^\m{}_\nu +V_\la\dl^\m_\nu,
\ee
i.e., has the so-called projective freedom.

\section{Goldstone gravitational field}

We consider the group bundle $Q\to X$ associated with $LX$ whose
typical fiber is the group $GL_4$ acting on itself via the adjoint
representation. It is equipped with the coordinates
$(x^\la, q^\la{}_\m)$ as a subbundle of the tensor bundle $TX\ot
T^*X$ and admits a canonical section $q_0(x)=\dr_\m\ot dx^\m$.
The canonical left action $Q$ on any bundle associated with $LX$ is
given. In particular, action (9) on the tetrad bundle $\Si$ has the form
\be
\rho:(x^\la,q^\la{}_\m, \si^\m_a)\mapsto 
(x^\la,q^\la{}_\m \si^\m_a).
\ee
We now fix the tetrad field $h$ and let $\Ker_h\rho$ denote the set of
those elements of the group bundle $Q$ which leave $h$ invariant. This
is a subbundle of the bundle $Q$. We let $\Si_h$ denote the quotient of
the bundle $Q$ with respect to 
$\Ker_h\rho$. This bundle with the coordinates $\wt\si^\m_a=q^\m{}_\nu
h^\nu_a$ is isomorphic to the bundle $\Si$ equipped with the Lorentz
structure of the bundle associated with 
$L^hX$. As a result, we can define [29], [32] the representation 
\ben
&& \g_Q:(\Si_h\times T^*X)\op\ot_{\Si_h} (\Si_h\times S^h) \to (\Si_h\times
S^h), \nonumber \\
&&\g_Q: (\wt\si,dx^\m)\mapsto q^\m{}_\nu h^\nu_a\g^a= \wt\si^\m_a\g^a.
\label{a12} 
\een
Therefore, we can interpret the section $\wt h \neq h$ of the bundle
$\Si_h$ as an effective tetrad field and  
$\wt g^{\m\nu}=\wt h^\m_a \wt h^\nu_b\eta_{ab}$ as an effective metric.
We note that $\wt h$ is not a new tetrad field and $\wt g$ is not a new
metric, because the covectors $\wt h^a=\wt h^a_\m dx^\m$ are realized
by $\g$-matrices in the same representation as the covectors
$h^a=h^a_\m dx^\m$ and Greek indices are raised and lowered by the
background metric
$g^{\m\nu}=h^\m_a h^\nu_b\eta_{ab}$.

We thus obtain a variant of the relativistic theory of gravity (RTG)
[36], [37] in the case of a background tetrad field $h$ and a dynamic
gravitational field $q$, which can be inerpreted as a Goldstone field
in the framework of gauge theory. We construct the Dirac operator and
the full Lagrangian of the gauge theory with a Goldstone gravitational
field. 

We consider the spinor bundle $S^h$ associated with the backgrounf
tetrad field $h$, and spinor connection $K_h$ given by (20) on 
$S^h\to X$ induced by the linear connection $K$ given by (12) on $X$.
Using the covariant differential $D$ (see (21)) determined by this
connection and the representation $\g_Q$ given by (27), we can
construct the Dirac operator 
\mar{a13}\beq
\Delta_Q= q^\la{}_\m h^\m_a\g^a D_\la. \label{a13}
\eeq
Under restriction to the canonical section $q_0(X)$, Dirac operator
(28) reduces to the Dirac operator $\Delta_h$ given by (22) on $S^h$
for fermionic fields in the presence of a background tetrad field 
$h$ and a linear connection $K$. 

We thus obtain an extension of the RTG where the Godstone gravitational
field $q$, the general linear connections $K$, and the Dirac fermionic
fields in the presence of a background field $h$ appear as dynamic variables.
The manifold of jets $J^1Z$ of the bundle
\be
Z=Q\times C\times S^h,
\ee
parametrized by the coordinates $(x^\la,q^\m{}_\nu,
k_\la{}^\m{}_\nu,y^A)$ serve as the configuration space of such a
model. The full Lagrangian on this configuration space can be chosen in
the form of a sum
\be
L=L_{AM} +L_q(q,g) +L_D  
\ee
of the Lagrangian $L_{AM}$ of the affine-metric theory expressed via
the components of curvature tensor (26) contracted using the effective
metric $\wt\si^{\m\nu}=\wt\si^\m_a\wt\si^\nu_b\eta^{ab}$, 
the Lagrangian $L_q$ of a Goldstone gravitational field $q$ in which
contraction is performed using the background metric $g$, and the
Dirac Lagrangian $L_D$ given by (25) in which
the tetrad gravitational field $\si$ is replaced with the effective
tetrad field $\wt \si$.
In particular, setting 
\be
L_{AM}=(-\la_1R+\la_2)\nm{\wt\si}^{-1/2},  \quad
L_q=\la_3g_{\m\nu}\wt\si^{\m\nu}\nm{\si}^{-1/2}, \quad L_D=0, 
\ee
where $R=\wt \si^{\m\nu}R^\al{}_{\m\al\nu}$, we recover the usual
Lagrangian of the RTG.

\end{document}